# Frequency Locking Via Phase Mapping Of Remote Clocks Using Quantum Entanglement

M.S. Shahriar[1,2]

[1]Dept. of Electrical and Computer Engineering, Northwestern University, Evanston, IL 60208

[2]Research Laboratory of Electronics, Massachusetts Institute of Technology, Cambridge, MA 02139

**Abstract**

Recently, we have shown how the phase of an electromagnetic field can be determined by measuring the *population* of either of the two states of a two-level atomic system excited by this field, via the so-called Bloch-Siegert oscillation resulting from the interference between the co- and counter-rotating excitations. Here, we show how a degenerate entanglement, created without transmitting any timing signal, can be used to teleport this phase information. This phase-teleportation process may be applied to achieve frequency-locking of remote oscillators, thereby facilitating the process of synchronizing distant clocks.

The task of synchronizing a pair of clocks that are separated in space is important for many practical applications, such as the global positioning system (GPS)[1] and the very large base interferometry (VLBI).[2] Conventionally, the synchronization is performed by transmitting timing signals between the clocks. Consider first the *ideal* situation where the intervening medium is stable and fully characterized. Assume furthermore that the special and general relativistic corrections can be determined and applied properly. The accuracy of the synchronization process is then limited by the uncertainty in the timing signal. The best result achievable is limited by the signal to noise ratio (SNR). Given enough resources, one can eliminate sources of systematic noises, so that the fundamental constraint is the shot noise limit (SNL), which is a manifestation of the Heisenberg uncertainty principle of quantum mechanics. This situation is typical of virtually all processes in metrology. In principle, specially prepared quantum states can reduce the effective noise below the SNL. However, the exotic nature of these states imposes severe constraints on the signal level, so that in practice the actual SNR achievable this way is far below what can be achieved using classical states. Most of the recent proposals[3-5] for achieving improved clock synchronization using quantum processes suffer from the same constraint, so that in practice they are inferior to classical approaches. Thus, given the current state of technology, quantum mechanical effects is not likely to help in the process of clock synchronization under the *ideal* situation.

Consider next the *realistic* situation where the density of the intervening medium fluctuates randomly, leading to a corresponding fluctuation in the time needed for a signal to travel between the clocks. Under this condition, it is *fundamentally* impossible, barring superluminal propagation, to synchronize the clocks exactly. This follows from the principle of special relativity, which is built on the axiom that there exists a maximum speed --- namely, the speed of light in vacuum --- at which information can propagate. As such, the notion of clock synchrony is *defined* with respect to the time it takes for light to traverse the distance between the clocks. It then follows that if this travel time itself is fluctuating, then the clock synchrony is undefined, and cannot be achieved on the timescale of the fluctuation. One can define and establish only an *average* synchrony, valid only for timescales longer than that of the fluctuation. In all situations of practical interest, clock synchronization always implies the achievement of this *average synchrony*. We might wonder whether the so-called non-local quantum correlation realizable via entanglement between particles that are separated in space might enable one to circumvent this limitation. This potential promise has been the key reason for all the interest in quantum clock synchronization. However, since quantum mechanics is reconciled with special relativity, such an entanglement does not allow superluminal propagation of information. By the same token, it cannot circumvent the path-length fluctuation related limitation on clock synchronization.[6] The clock synchronization proposal of Jozsa et al[7] is constrained fundamentally by the same reason.

An alternative way to improve the average synchrony is through frequency locking. Specifically, consider a typical application where each of the timekeepers in question is an atomic clock (rubidium or cesium), which can be modeled as a frequency counter referenced to a master oscillator (MO), which in turn is locked to a metastable atomic transition. Most of the recent proposals about clock synchronization, including the Jozsa protocol[7], make the assumption that the MO in each clock continues to operate at some ideal transition frequency. In practice, however, this is not the case. The MO frequency for each clock undergoes shifts and drifts due to a host of reasons. These fluctuations lie at the heart of clock asynchrony. As such, minimizing the relative drifts in the MO frequencies is perhaps the most effective way to minimize the error in clock synchronization. This approach opens up new possibilities for exploring whether quantum mechanical effects may outperform classical approaches. In this paper, we propose a new technique for locking the frequencies of two distance oscillators, via the process of *wavelength teleportation.*

In this technique, the phase variation of an oscillator is first mapped by Alice (keeper of the first clock) to the wave-functions of an array of atoms, via the use of the Bloch-Siegert oscillation, which results from an interference between the co- and counter-rotating parts of a two-level excitation.[8] The maximum number of atoms needed to encode the phase variation can be very small, and is given by the Nyquist sampling criterion. Distant entanglement, produced using an asynchronous technique[9], is used to teleport the quantum state of each of these atoms to a matching atom with Bob (keeper of the second clock). Bob can thus recreate the exact phase variation of Alice's MO locally, and compare with the same for his MO. We discuss the potential constraints and advantages of this approach after presenting the scheme in detail.

Consider first a situation where Alice and Bob each has an atom that has two degenerate ground states ($|1>$ and $|2>$), each of which is coupled to a higher energy state ($|3>$), as shown in figure 1. We assume the 1-3 and 2-3 transitions are magnetic dipolar, and orthogonal to each other, with a transition frequency $\omega$. For example, in the case of $^{87}$Rb, $|1>$ and $|2>$ correspond to $5^2P_{1/2}:|F=1,m_F=-1>$ and $5^2P_{1/2}:|F=1,m_F=1>$ magnetic sublevels, respectively, and $|3>$ corresponds to $5^2P_{1/2}:|F=2,m_F=0>$ magnetic sublevel[10]. Left and right circularly polarized magnetic fields, perpendicular to the quantization axis, are used to excite the 1-3 and 2-3 transitions, respectively. The rubidium clock is typically stabilized with respect to the $5^2P_{1/2}:|F=1,m_F=0>$ to $5^2P_{1/2}:|F=2,m_F=0>$ transition, which is excited by a magnetic field parallel to the quantization axis. We take $\omega$ to be the same as the clock frequency $\omega_c$.

We assume that Alice and Bob's fields at $\omega$ have the form $B_A=B_{ao}Cos(\omega t+\phi)$ and $B_B=B_{bo}Cos(\omega t+\chi)$, respectively. The origin of the time variable, $t$, is therefore arbitrary, and does not affect the phase difference, $\Omega \equiv (\phi-\chi)$. The clocks are assumed to be in phase if $\Omega=0$, so that if Bob determines that at some instant his magnetic field is maximum and positive in some direction $r_b$, then Alice will also find her magnetic field to be maximum and positive in some direction $r_a$ at the same instant. As long as Alice and Bob agree on this definition of phase-locking, and use the same definitions all the time, then $r_b$ and $r_a$ do not have to be the same. During the magnetic resonance excitations, the value of any dc magnetic field will be assumed to be vanishing. Symmetry then dictates that any physical observable will be independent of the choice of the quantization axis, as long as it is perpendicular to $r_a$ for Alice, and perpendicular to $r_b$ for Bob. In order to describe our protocol, we now summarize briefly the theory behind the Bloch-Siegert oscillation that occurs when a two-level interaction is considered without the rotating wave approximation (RWA)[11-14], and presented in greater detail in ref. 8. We also describe the condition for the time reversal of an arbitrary evolution under this condition, another necessary element of our protocol.

Consider, for example, the excitation of the $|1>_A \leftrightarrow |3>_A$ transition. In the dipole approximation, the Hamiltonian can be written as:

$$\hat{H} = \begin{bmatrix} 0 & g(t) \\ g(t) & \varepsilon \end{bmatrix} \qquad \ldots(1)$$

and the state vector is written as:

$$|\xi(t)\rangle = \begin{bmatrix} C_{1A} \\ C_{3A} \end{bmatrix} \quad \ldots(2)$$

where g(t) = -g$_o$[exp(iωt+iφ)+c.c.]/2, and we assume that ε=ω corresponding to resonant excitation. Here, we have assumed that the polarization of Alice's field can be changed to excite either |1>$_A$↔|3>$_A$ or the |2>$_A$↔|3>$_A$ transition, in a mutually exclusive manner. We now perform a rotating wave transformation by operating on |ξ(t)> with the unitary operator Q, given by:

$$\hat{Q} = \begin{bmatrix} 1 & 0 \\ 0 & \exp(i\omega t + i\phi) \end{bmatrix} \quad \ldots(3)$$

The Schroedinger equation then takes the form (setting ℏ=1):

$$\frac{\partial |\tilde{\xi}(t)>}{\partial t} = -i\tilde{H}(t)|\tilde{\xi}(t)> \quad \ldots(4)$$

where the effective Hamiltonian is given by:

$$\tilde{H} = \begin{bmatrix} 0 & \alpha(t) \\ \alpha^*(t) & 0 \end{bmatrix} \quad \ldots(5)$$

with α(t)= -g$_o$[exp(-i2ωt-i2φ)+1]/2, and the rotating frame state vector is:

$$|\tilde{\xi}(t)> \equiv \hat{Q}|\tilde{\xi}(t)> = \begin{bmatrix} \tilde{C}_{1A} \\ \tilde{C}_{3A} \end{bmatrix} \quad \ldots(6)$$

Given the periodic nature of the effective Hamiltonian, the general solution to eqn. 5 can be written as:

$$|\tilde{\xi}(t)> = \sum_{n=-\infty}^{\infty} |\xi_n\rangle \beta^n \quad \ldots(7)$$

where β=exp(-i2ωt-i2φ), and

$$|\xi_n\rangle \equiv \begin{bmatrix} a_n \\ b_n \end{bmatrix} \quad \ldots(8)$$

Inserting eqn. 7 in eqn. 4, and equating coefficients with same frequencies, we get, for all n:

$$\dot{a}_n = i2n\omega a_n + ig_o(b_n + b_{n-1})/2$$
$$\dot{b}_n = i2n\omega b_n + ig_o(a_n + a_{n+1})/2 \quad \ldots(9)$$

Figure 2 shows a pictorial representation of these equations. Here, the coupling between a$_o$ and b$_o$ is the conventional one present when the RWA is made. The coupling to additional levels results from virtual multiphoton processes in the absence of the RWA. The couplings to the nearest neighbors, a$_{\pm 1}$ and b$_{\pm 1}$ are detuned by an amount 2ω, and so on. To the lowest order in (g$_o$/ω), we can ignore terms with |n|>1, thus yielding a truncated set of six eqns.:

$$\dot{a}_o = ig_o(b_o + b_{-1})/2 \quad (10.1)$$
$$\dot{b}_o = ig_o(a_o + a_1)/2 \quad (10.2)$$

$$\overset{\circ}{a}_1 = i2\omega a_1 + ig_o(b_1 + b_o)/2 \qquad (10.3)$$

$$\overset{\circ}{b}_1 = i2\omega b_1 + ig_o a_1/2 \qquad (10.4)$$

$$\overset{\circ}{a}_{-1} = -i2\omega a_{-1} + ig_o b_{-1}/2 \qquad (10.5)$$

$$\overset{\circ}{b}_{-1} = -i2\omega b_{-1} + ig_o(a_{-1} + a_o)/2 \qquad (10.6)$$

In order to solve these equations, one may employ the method of adiabatic elimination valid to first order in $\sigma \equiv (g_o/4\omega)$. To see how this can be done, consider first the last two equations: 10.5 and 10.6. In order to simplify these two equations further, one needs to diagonalize the interaction between $a_{-1}$ and $b_{-1}$. To this end, we define $\mu_- \equiv (a_{-1} - b_{-1})$ and $\mu_+ \equiv (a_{-1} + b_{-1})$, which can be used to re-express these two equations in a symmetric form as:

$$\overset{\circ}{\mu}_- = -i(2\omega + g_o/2)\mu_- - ig_o a_o/2 \qquad (11)$$

$$\overset{\circ}{\mu}_+ = -i(2\omega - g_o/2)\mu_+ + ig_o a_o/2 \qquad (12)$$

Adiabatic following then yields (again, to lowest order in $\sigma$):

$$\mu_- \approx -\sigma a_o; \qquad \mu_+ \approx \sigma a_o \qquad (13)$$

which in turn yields:

$$a_{-1} \approx 0; \qquad b_{-1} \approx \sigma a_o \qquad (14)$$

In the same manner, we can solve eqns. 10.3 and 10.4, yielding:

$$a_1 \approx -\sigma b_o; \qquad b_1 \approx 0 \qquad (15)$$

Note that the amplitudes of $a_{-1}$ and $b_1$ are vanishing (each proportional to $\sigma^2$) to lowest order in $\sigma$, thereby justifying our truncation of the infinite set of relations in eqn. 9. Using eqns. 14 and 15 in eqns. 10.1 and 10.2, we get:

$$\overset{\circ}{a}_o = ig_o b_o/2 + i\Delta a_o/2 \qquad (16)$$

$$\overset{\circ}{b}_o = ig_o a_o/2 - i\Delta b_o/2 \qquad (17)$$

where $\Delta = g_o^2/4\omega$ is essentially the Bloch-Siegert shift. Eqns. 16 and 17 can be thought of as a two-level system excited by a field detuned by $\Delta$. With the initial condition of all the population in $|1\rangle_A$ at t=0, the only non-vanishing (to lowest order in $\sigma$) terms in the solution of eqn. 9 are:

$$a_o(t) \approx Cos(g_o t/2); \quad b_o(t) \approx iSin(g_o t/2)$$
$$a_1(t) \approx -i\sigma Sin(g_o t/2); \quad b_{-1}(t) \approx \sigma Cos(g_o t/2) \qquad (18)$$

Inserting this solution in eqn. 7, and reversing the rotating wave transformation, we get the following expressions for the components of eqn. 2:

$$C_{1A}(t) = Cos(g_o t/2) - 2\sigma\Sigma \cdot Sin(g_o t/2)$$
$$C_{3A}(t) = ie^{-i(\omega t + \phi)}[Sin(g_o t/2) + 2\sigma\Sigma^* \cdot Cos(g_o t/2)] \qquad (19)$$

where we have defined $\Sigma \equiv (i/2)\exp[-i(2\omega t + 2\phi)]$. To lowest order in $\sigma$ this solution is normalized at all times. Note that if Alice were to carry this excitation on an ensemble of atoms through for a $\pi/2$ pulse, and measure the population of the state $|1\rangle_A$ immediately (at t=$\tau$, the moment when the $\pi/2$ excitation ends), the result would be a signal given by

[1+2σSin(2ωτ+2φ)]/2, which contains information related to the amplitude and phase of her field.

Next, we consider the issue of exact time reversal of such an excitation. The Schroedinger eqn. (4) has the formal solution:

$$|\tilde{\xi}(t_2)> = \exp(-i\int_{t_1}^{t_2} \tilde{H}(t')dt')|\tilde{\xi}(t_1)> \quad \ldots(20)$$

If the RWA is made, then $\tilde{H}$ is time independent. In that case, if one starts an evolution at $t_1$, proceed for *any* duration T, then reverses the sign of $\tilde{H}$ by shifting the phase of the magnetic field by π, and continues with the evolution for another duration T, then the system returns back to the starting state. Of course, this can be verified explicitly using the well known solution of Rabi flopping. Here, however, RWA is not made, so that $\tilde{H}$ depends on time. Therefore, the exact reversal can be achieved in this manner only if T=mπ/ω for any integer value of m. Parenthetically, note that while we have considered direct excitations of the two-level systems, all the results derived above apply equally to the case where an off-resonant Raman excitation is used to couple the two levels[15-18].

Returning to the task at hand, our protocol starts by using a scheme, developed earlier by us[9] (note that this scheme works for $^{87}$Rb, for the choice of |1> and |2> as indicated above) to produce a degenerate entanglement of the form |ψ>=(|1>$_A$|2>$_B$ - |2>$_A$|1>$_B$)/√2. Next, Alice attenuates her field so that the counter-rotating term in the Hamiltonian can be ignored (this assumption is not essential for our conclusion, but merely simplifies the algebra somewhat), and excites a π-pulse coupling |2>$_A$ to |3>$_A$, and then stops the excitation. Similarly, Bob uses a field, attenuated as above, to excite a π-pulse coupling |2>$_B$ to |3>$_B$, and then stops the excitation. Using digital communications over a classical channel, Alice and Bob wait until they both know that these excitations have been completed. The resulting state is then given by :

|ψ(t)>=[|1>$_A$|3>$_B$exp(-iωt-iχ) - |3>$_A$|1>$_B$exp(-iωt-iφ)]/√2.   …(21)

The next step is for Alice to make a measurement along the |1>$_A$↔|3>$_A$ transition. For this process, she chooses a much larger value of $g_o$, so that the RWA can not be made. The state she wants to measure is *the one that would result if one were to start from state |1>$_A$, and evolve the system for a π/2 pulse using this stronger $g_o$*:

$$|+\rangle_A \equiv \frac{1}{\sqrt{2}}\left[\{1-2\sigma\Sigma\}|1\rangle_A + ie^{-i(\omega t+\phi)}\{1+2\sigma\Sigma^*\}|3\rangle_A\right] \quad \ldots(22)$$

where we have made use of eqn. 19. The state orthogonal to |+>$_A$ results from a 3π/2 pulse:

$$|-\rangle_A \equiv \frac{1}{\sqrt{2}}\left[\{1+2\sigma\Sigma\}|1\rangle_A - ie^{-i(\omega t+\phi)}\{1-2\sigma\Sigma^*\}|3\rangle_A\right] \quad \ldots(23)$$

(Equivalently, the state of eqn. 23 results from a π/2 pulse excitation starting from -i|3>$_A$ ). To first order in σ, these two states are each normalized, and orthogonal to each other. As such, one can re-express the state of the two atoms in eqn. 21 as:

$$|\psi(t)\rangle = \frac{1}{\sqrt{2}}\left[|+\rangle_A|-\rangle_B - |-\rangle_A|+\rangle_B\right] \quad \ldots(24)$$

where we have defined:

$$|+\rangle_B \equiv \frac{1}{\sqrt{2}}\left[\{1-2\sigma\Sigma\}|1\rangle_B + ie^{-i(\omega t+\chi)}\{1+2\sigma\Sigma^*\}|3\rangle_B\right] \quad \ldots(25)$$

$$|-\rangle_B \equiv \frac{1}{\sqrt{2}}\left[\{1+2\sigma\Sigma\}|1\rangle_B - ie^{-i(\omega t+\chi)}\{1-2\sigma\Sigma^*\}|3\rangle_B\right] \quad \ldots(26)$$

She can measure the state $|+\rangle_A$ by taking the following steps: (i) Shift the phase of the B-field by $\pi$, (ii) Fine tune the value of $g_o$ so that $g_o=\omega/2m$, for an integer value of m, (iii) apply the field for a duration of $T=\pi/2g_o$, and (iv) detect state $|1\rangle_A$. Note that the constraint on $g_o$ ensures that $T=m\pi/\omega$, which is necessary for time reversal to work in the absence of the RWA. Once Alice performs this measurement, the state for Bob collapses to $|-\rangle_B$, given in eqn. 26. Note that if σ is neglected, then the measurement produces a $|-\rangle_B$ that contains no information about the phase of Alice's clock, which is analogous to the Jozsa protocol[7].

In the present case, $|-\rangle_B$ does contain information about the amplitude and the phase of Alice's clock signal. In order to decipher this, Bob measures his state $|1\rangle_B$. The probability of success is:

$$p_\phi \equiv |{}_B\langle 1|-\rangle_B|^2 = \frac{1}{2}[1+2\sigma Sin(2\phi)]. \quad \ldots(27)$$

where we have kept terms only to the lowest order in σ. Of course, the value of φ(mod 2π), the phase difference, can not be determined from knowing Sin(2φ) alone. However, this whole process can be repeated after, for example, Alice shifts the phase of her B-field by π/2, so that Bob can determine the value of Cos(2φ). It is then possible to determine the value of φ (mod 2π) unambiguously.

The overall process can be carried out in one of two ways. First, consider the situation where Alice and Bob starts with X pairs of atoms, and entangle each pair in the form of equation 24. Then, over a digital communication channel, Alice sends Bob a list of the M atoms she found in state $|1\rangle_A$ after performing her measurement process described above. Bob performs his measurement only on this subset of atoms. Suppose he finds L number of atoms in state $|1\rangle_B$. Then:

$$\eta \equiv \left(\frac{L}{M}-\frac{1}{2}\right) \to \sigma Sin(2\phi) \quad , \textit{for large M} \quad \ldots(28)$$

Thus, the value of η determined asymptotically for a large number of entangled pairs will reveal the value of Sin(2φ). Alternatively, if only a single pair of atoms is available, then the same result can be obtained by repeating the whole process X times, assuming that φ remains unchanged during the time needed for the process.

Note that what is determined by Bob is φ, not Ω. Thus, it is not possible to measure the absolute phase difference in this manner. In fact, one must transmit a timing signal in order to determine Ω. This process is potentially hampered by the presence of undetermined fluctuations in the intervening pathlength. As discussed in the introduction, we have proven that in general quantum entanglement offers no advantage over a classical approach in determining Ω in the presence of such an undetermined source of noise. However, one could use this approach of phase teleportation in order to achieve frequency locking of two remote oscillators. One could use this approach of phase teleportation in order to achieve frequency locking of two remote oscillators. This is illustrated in figure 3.

Briefly, assume that Bob has an array of N atoms. Assume further that Alice also has an identical array of atoms. *For our protocol, the physical separations between the neighboring*

*atoms do not have to match.* In principle, one can create such an identical pair of arrays by embedding N rows of atoms (or quantum dots) in a substrate patterned lithographically, with two atoms in each row, and then splitting it in two halves. To start with, the corresponding atoms in each array are entangled with each other using the asynchronous approach of ref. 9. Here, we assume that the two clocks may differ in frequency. The frequency-locking algorithm then proceeds as follows. Alice and Bob both apply their fields parallel to their arrays of atoms, so that the phase variation is $2\pi$ over their respective wavelengths. After Alice makes her measurements of the state $|+>_A$, using the same set of steps as described above, she informs Bob, over a classical communication channel, the indices of her atoms that were found in this state. Bob now measures the state $|->_B$ for this subgroup of atoms only, using an analogous set of time-reversed excitation steps which ends in observing his atom in state $|3>_B$. For a given atom in this subgroup, the phase of his field at that location at the time Bob starts the measurement affects the probability of success in finding the atom in state $|3>_B$ at the end of the measurement process. This phase is varied as Bob repeats the measurement different measurement-starting-times (modulo $2\pi/\omega_B$, where $\omega_B$ is the frequency of Bob's clock). It is easy to show that there exists a choice of this phase for which the probability of success is 100%. However, the success probability for atoms (in the post-selection subgroup) would vary with location if the frequencies of Bob's and Alice's clocks are not the same. This effect can be used by Bob to adjust his clock frequency, thereby achieving frequency locking. The Nyquist sampling criterion dictates that the number of atoms in this subgroup can be as low as only two, so that N can be quite small, thus making this protocol potentially practicable.

To summarize, previously we have shown how the phase of an electromagnetic field can be determined by measuring the population of either of the two states of a two-level atomic system excited by this field, via the so-called Bloch-Siegert oscillation resulting from the interference between the co- and counter-rotating excitations. Here, we show how a degenerate entanglement, created without transmitting any timing signal, can be used to teleport this phase information, thus making it possible to achieve frequency-locking of remote oscillators, thereby facilitating the process of synchronizing distant clocks.

This work was supported by DARPA grant # F30602-01-2-0546 under the QUIST program, ARO grant # DAAD19-001-0177 under the MURI program, and NRO grant # NRO-000-00-C-0158.

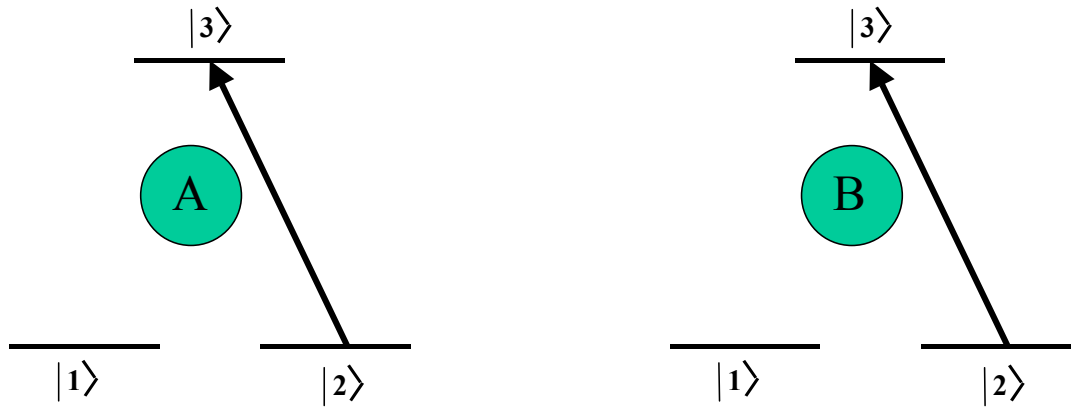

*Figure 1. Schematic illustration of the basic protocol for phase locking two remote clocks, one with Alice (A), and the other with Bob (B), without transmitting a clock signal directly. The model energy levels can be realized, for example, using the metastable hyperfine Zeeman sublevels of $^{87}$Rb atoms, as detailed in the text.*

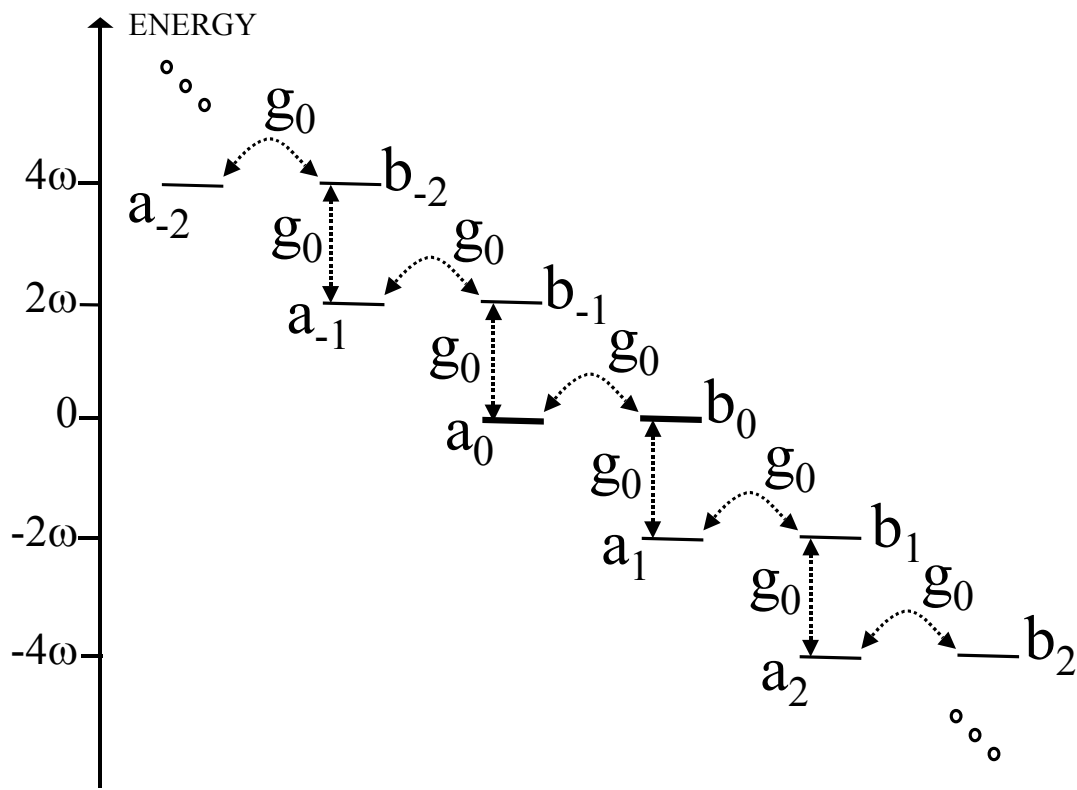

*Figure 2. Schematic illustration of the multiple orders of interaction when the rotating wave approximation is not made. The strengths of the first higher order interaction, for example, is weaker than the zeroth order interaction by the ratio of the Rabi frequency, $g_o$, and the effective detuning, $2\omega$.*

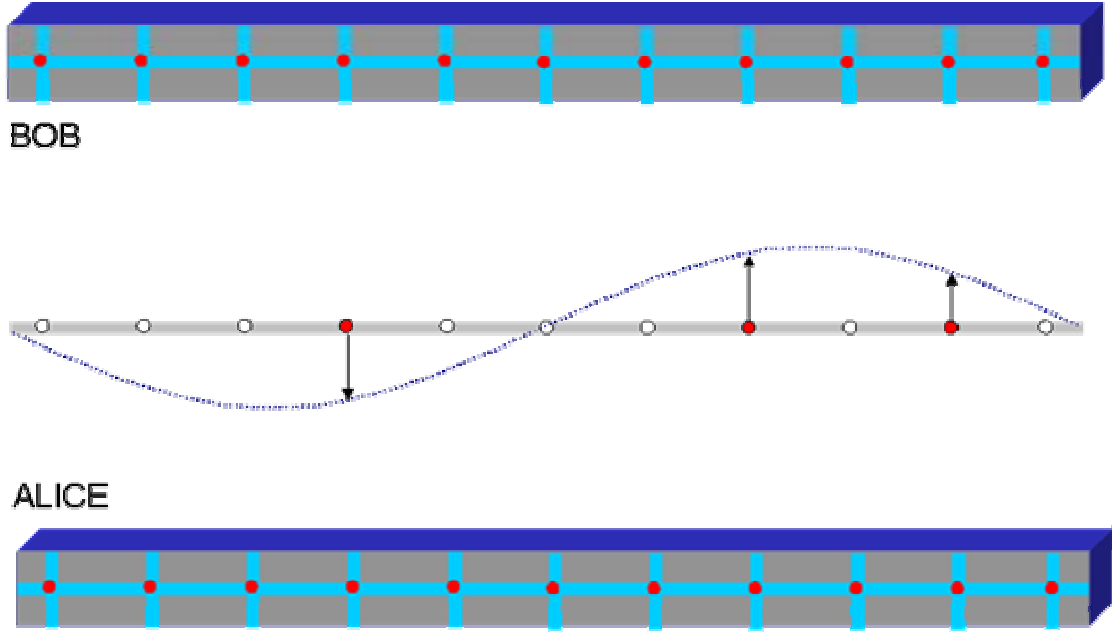

*Figure 3. Schematic illustration of the process to be employed for remote frequency locking. See text for details.*